\documentclass[prd,twocolumn,a4paper,superscriptaddress,floatfix]{revtex4}
\usepackage{amsmath,amssymb}
\usepackage{graphicx}

\begin{document}
\newcommand{\be}{\begin{equation}}
\newcommand{\ee}{\end{equation}}
\newcommand{\bq}{\begin{eqnarray}}
\newcommand{\eq}{\end{eqnarray}}
\newcommand{\n}{\nonumber}

\title{Solitonic description of interface profiles in competition models}

\author{T. Azevedo}
\email[Electronic address: ]{tiberio@dfte.ufrn.br}
\affiliation{Departamento de F\'isica Te\'orica e Experimental, Universidade Federal do Rio Grande do Norte, RN, Brazil}
\affiliation{Institute for Biodiversity and Ecosystem Dynamics, University of Amsterdam, Science Park 904, 1098 XH Amsterdam, The Netherlands}
\author{L. Losano}
\email[Electronic address: ]{losano@fisica.ufpb.br}
\affiliation{Departamento de F\'{\i}sica, Universidade Federal da Para\'{\i}ba 58051-970 Jo\~ao Pessoa, PB, Brazil}
\author{J. Menezes}  
\email[Electronic address: ]{jmenezes@ect.ufrn.br} 
\affiliation{Institute for Biodiversity and Ecosystem Dynamics, University of Amsterdam, Science Park 904, 1098 XH Amsterdam, The Netherlands}
\affiliation{Escola de Ci\^encias e Tecnologia, Universidade Federal do Rio Grande do Norte\\
Caixa Postal 1524, 59072-970, Natal, RN, Brazil}

\begin{abstract} 
We consider systems with two competing species whose actions are completely symmetric, with same mobility, reproduction and competition rates. 
Numerical implementations of the model in two and three-dimensional space show that regions of single species are formed by spontaneous symmetry breaking. We propose a theoretical formalism for describing the static profile of the interfaces of empty spaces separating domains with different species.
We compute the topological properties of the interfaces and show that these theoretical functions are useful to the understanding of the dynamics of the network. Finally, we compare the theoretical functions with results from the numerical implementation of the mean field equations and verify that our model fits well the properties of interfaces. 
\end{abstract}

\maketitle
\section{Introduction} 
The different ways the species interact each other is responsible for the large variety of biodiversity observed in Nature.
The arising and evolution of spatial patterns have proved to play an important role on the understanding of the dynamics of populations and ecosystems\cite{nowak06evolutionaryDynamicsBOOK,sole2006selforganization}. 

The generalized May-Leonard model (rock-paper-scissors game) has been an important tool to describe the interactions of competing species  \cite{May-Leonard,Kerr2002,Reichenbach2007,PhysRevLett.77.2125,PhysRevLett.99.238105}.  
Numerical implementations of the mean field equations show that the larger number of strategies the more complex spatial patterns are present. Furthermore, the spatial displayment of the individuals is directly related to the parameters which control their movement, predation and reproduction actions \cite{smith82,Szabó200797}. 
   
The rock-paper-scissors game has been applied successfully to describe the population dynamics in many cases, like communities of coral reef invertebrates \cite{coral} and lizards in the inner Coast Range of California \cite{lizards}. 
Moreover, experimental tests using microbial laboratory cultures of three strains of colicinogenic \textit{Escherichia coli} showed that
even though a cyclic dominance is present, the biodiversity is achieved only if local interactions are considered \cite{bacteria}.
As a result, stable spiral patterns of domains of single species were formed.
The same dynamics is present in more complex systems with a larger number of strategies \cite{PhysRevE.86.036112,Boerlijst199117}.

Furthermore, investigations of the simplest case of two competing species is useful to predict the conditions of persistence of species
\cite{Kirkup2004,1742-5468-2012-07-P07014,PhysRevE.87.032148,0305-4470-38-30-005,PhysRevE.64.042902,Lutz2013286}. 
The results of numerical simulations agree with experimental researches of systems of different species of butterflies \cite{but1,but2,but3}, for example. It has been shown in Ref. ~\cite{PhysRevE.86.031119,PhysRevE.86.036112,PhysRevE.89.042710,Avelino2014393} that the interfaces of empty sites, created by the competition between species, enter into a scaling regime where the characteristic scale of the network $L$ grows as $L \propto t^{1/2}$. This is a typical scaling law associated with the dynamics of the nonlinear systems \cite{PhysRevLett.98.145701}. In addition, the generation of strings networks of empty sites has been recently shown in Ref.~\cite{Avelino2014393}, where three-dimensional networks are considered. In this case, due to the specific symmetry in the competition rules, domains of single species are distributed around roughly circular areas where attacks mostly take place. These spatial pattern networks obeys the same scaling law of the linear interfaces and their dynamics can be identified with cosmic strings in Cosmology.

The spatial pattern networks have been studied by means of mathematical analysis and numerical implementation of the Lotka-Volterra equations \cite{Volterra,doi:10.1021/ja01453a010}. Provided that the dynamics is similar to domains wall networks in Physics, we claim that the description of the solitonic aspects of the interface may reveal interesting conclusions about the population dynamics.
In this paper, we focus on the simplest case with two species equal conditions of competition. Our main goal is to present a potential with $Z_2$ symmetry, which is spontaneously broken to generate the interfaces.
This new formalism allows the comparison of the population dynamics with topological defects, extensively studied in other scenarios in cosmology and condensed matter \cite{Stavans1989,Glazier1992,Flyvbjerg1993,Monnereau1998,Weaire2000,Kim2006,Avelino2008,Avelino2010,PhysRevLett.101.087204}.

In the next section, we present the mean field equations of the competition model with two species. 
We also highlight the role played by the vacancies for the formation and evolution of the interface networks. 
In Section III we introduce a new scalar field for describing the 
spontaneous symmetry breaking process. In Section IV, we find the interface profiles and compute all topological properties provided by theoretical framework, which are associated with the physical properties of the interfaces in Section V.
In Section VI we discuss our results by comparing the theoretical model with results of the numerical implementation of the mean field equations. Finally, the conclusions are presented in Section VII.  

\section{The interface networks} 
\begin{figure}
	\centering
	\includegraphics*[width=4.1cm]{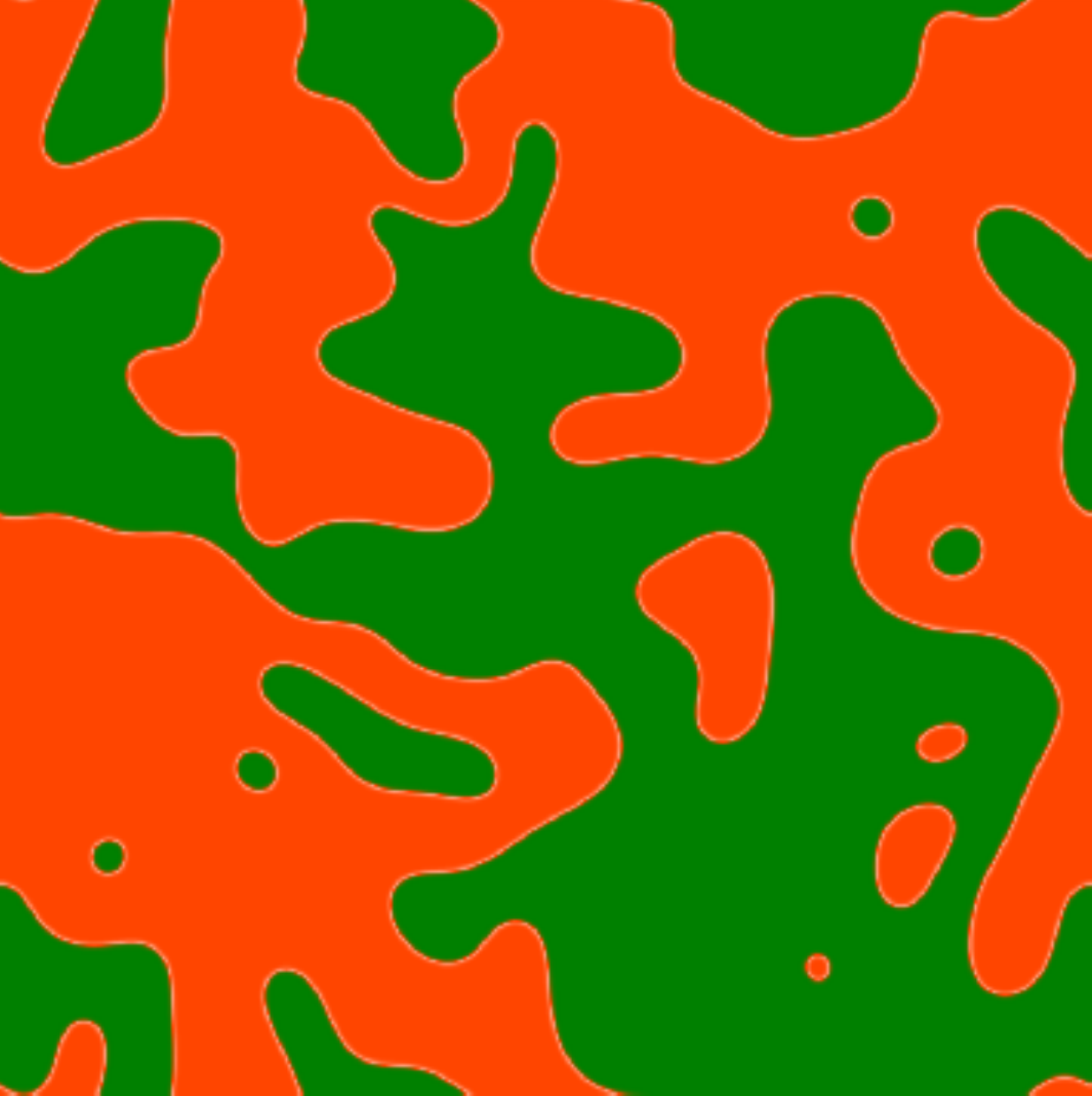}
	\includegraphics*[width=4.1cm]{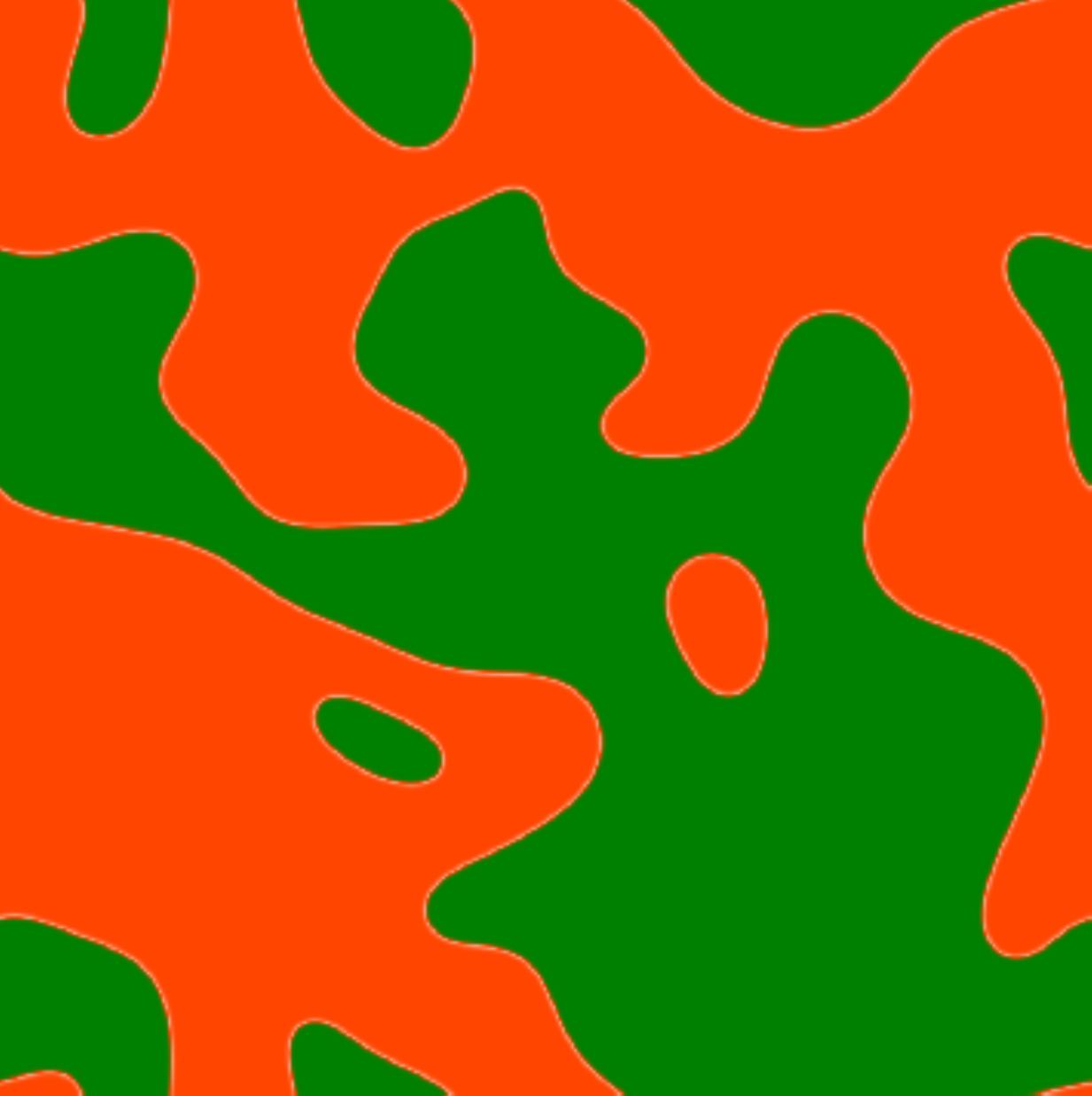}
	\includegraphics*[width=4.1cm]{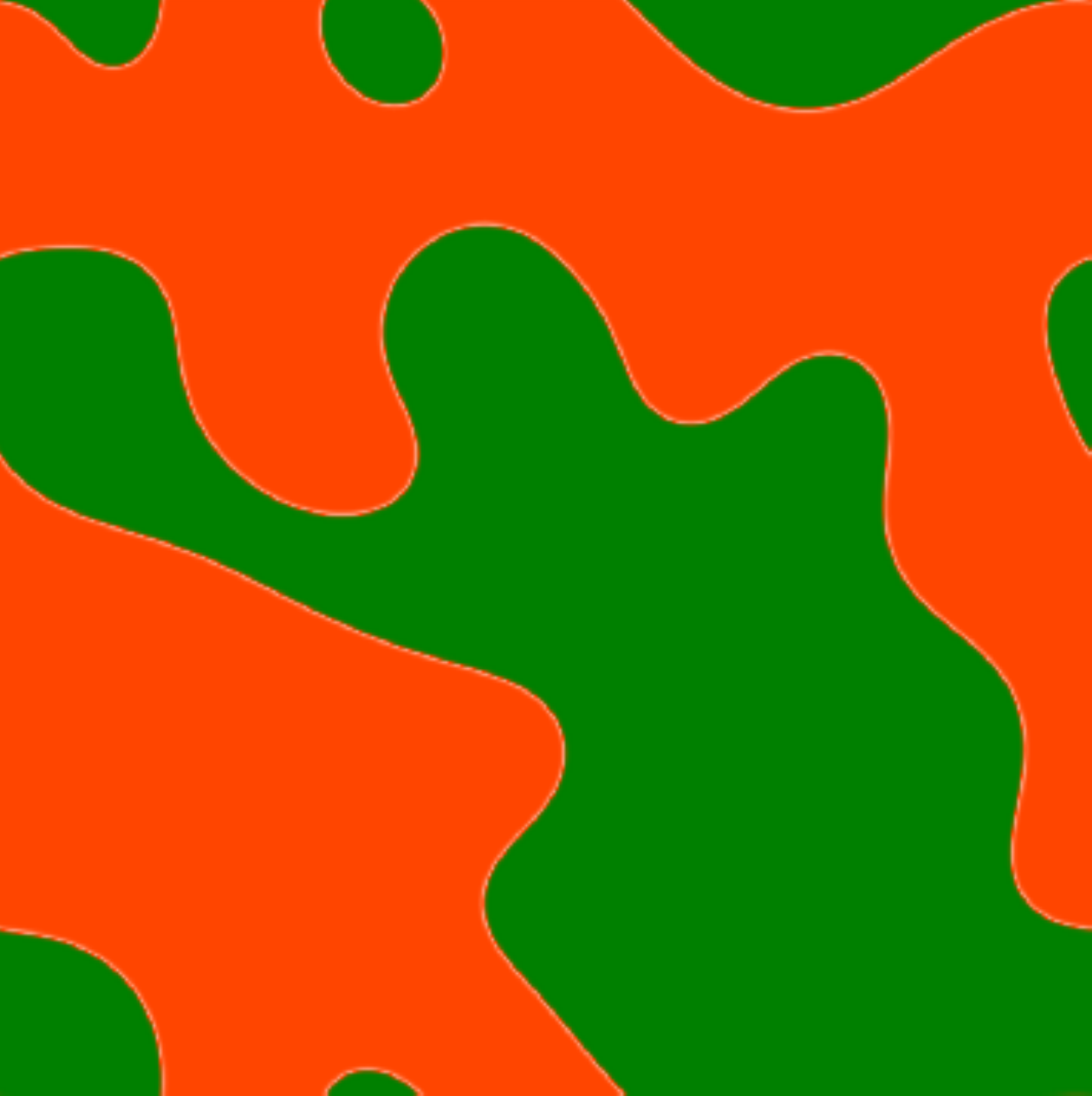}
	\includegraphics*[width=4.1cm]{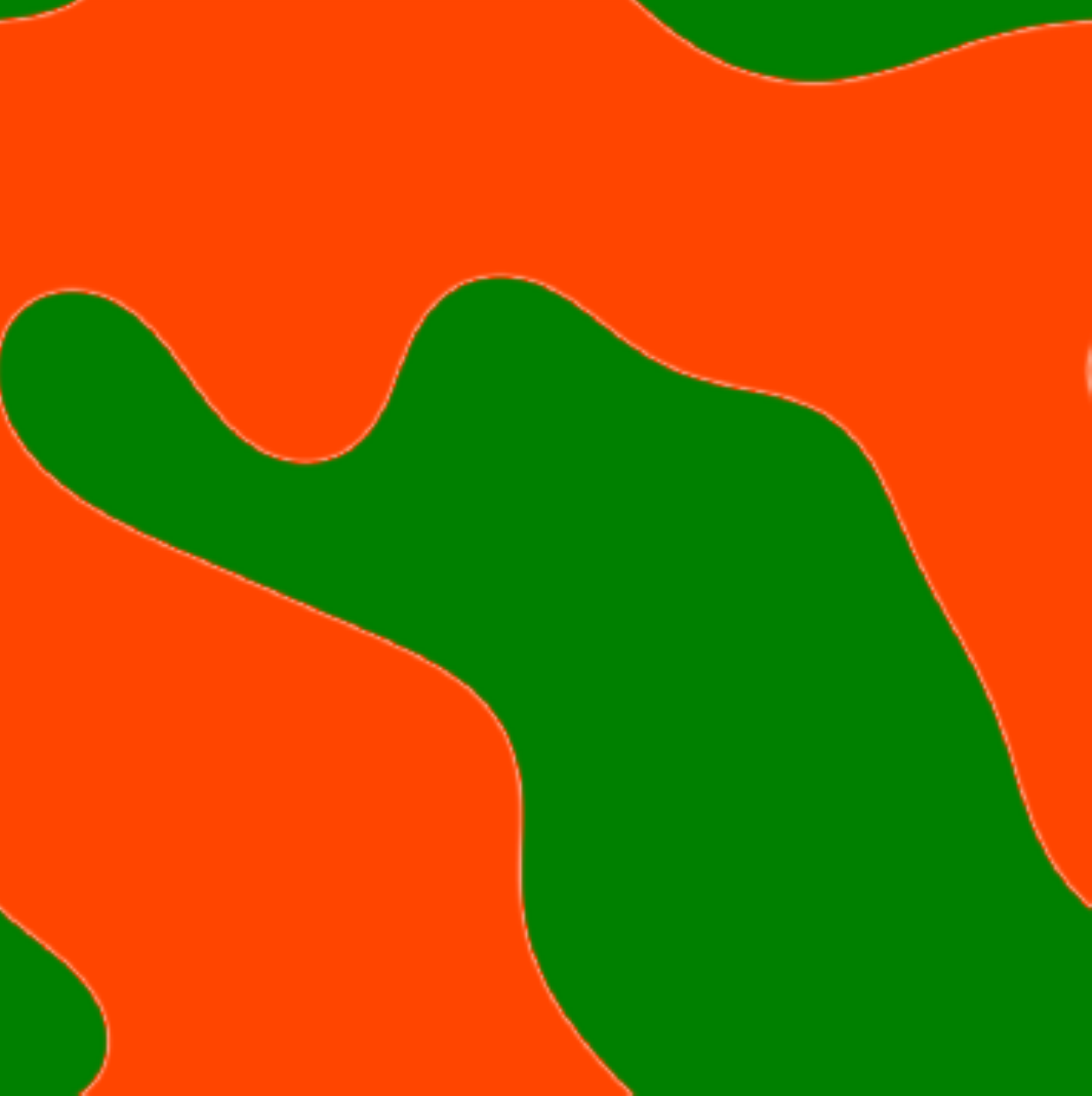}
\caption{(Color online). Snapshots of the implementation of the mean field simulations. The snapshots were 
taken after $1000$, $2000$, $4000$ and $8000$ generations, respectively. The colors orange and green represent
the domains where $\phi_1=1$ and $\phi_2=1$, respectively. }
	\label{fig3}
\end{figure}

\begin{figure}
	\centering
	\includegraphics*[width=4.1cm]{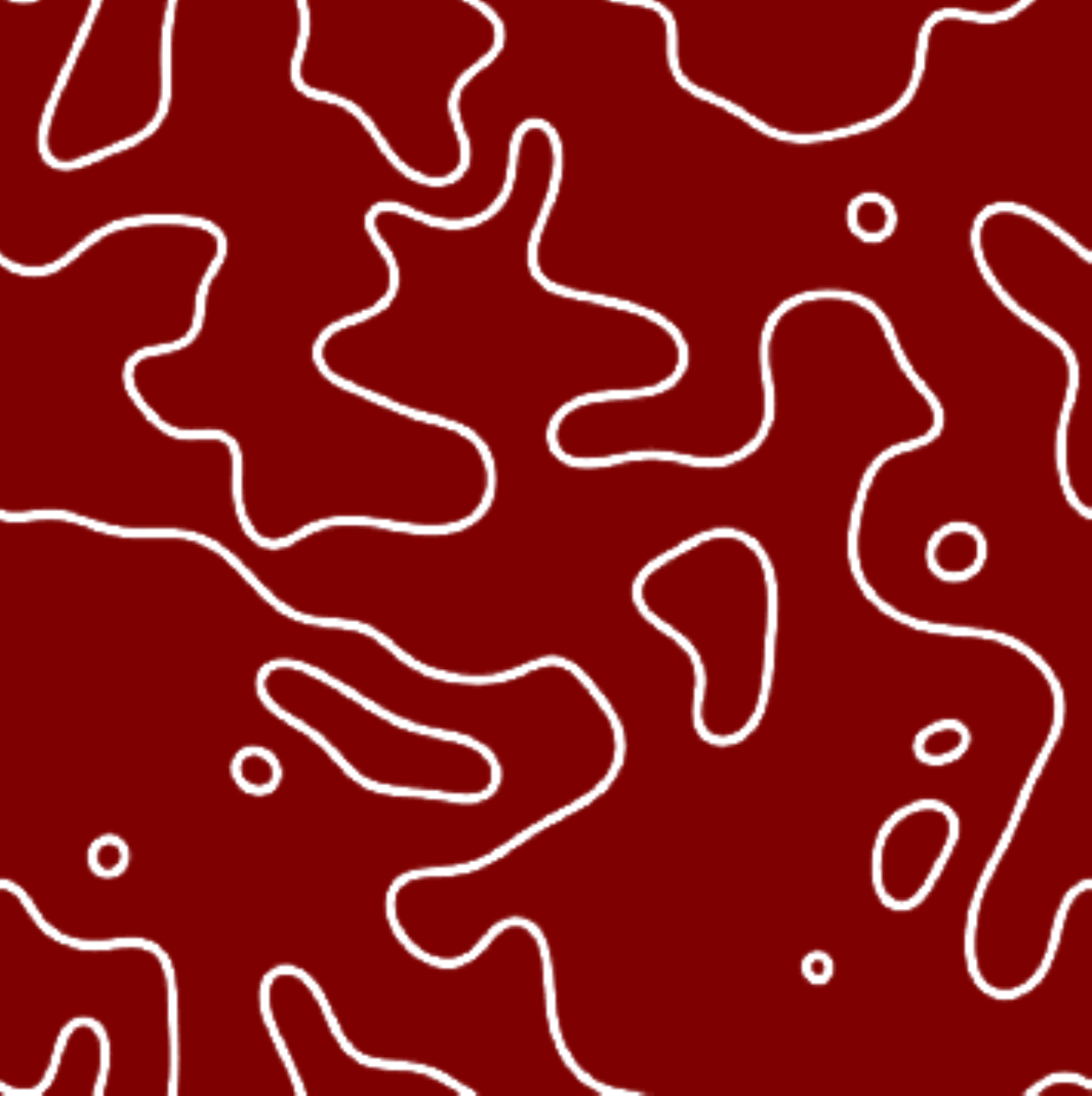}
	\includegraphics*[width=4.1cm]{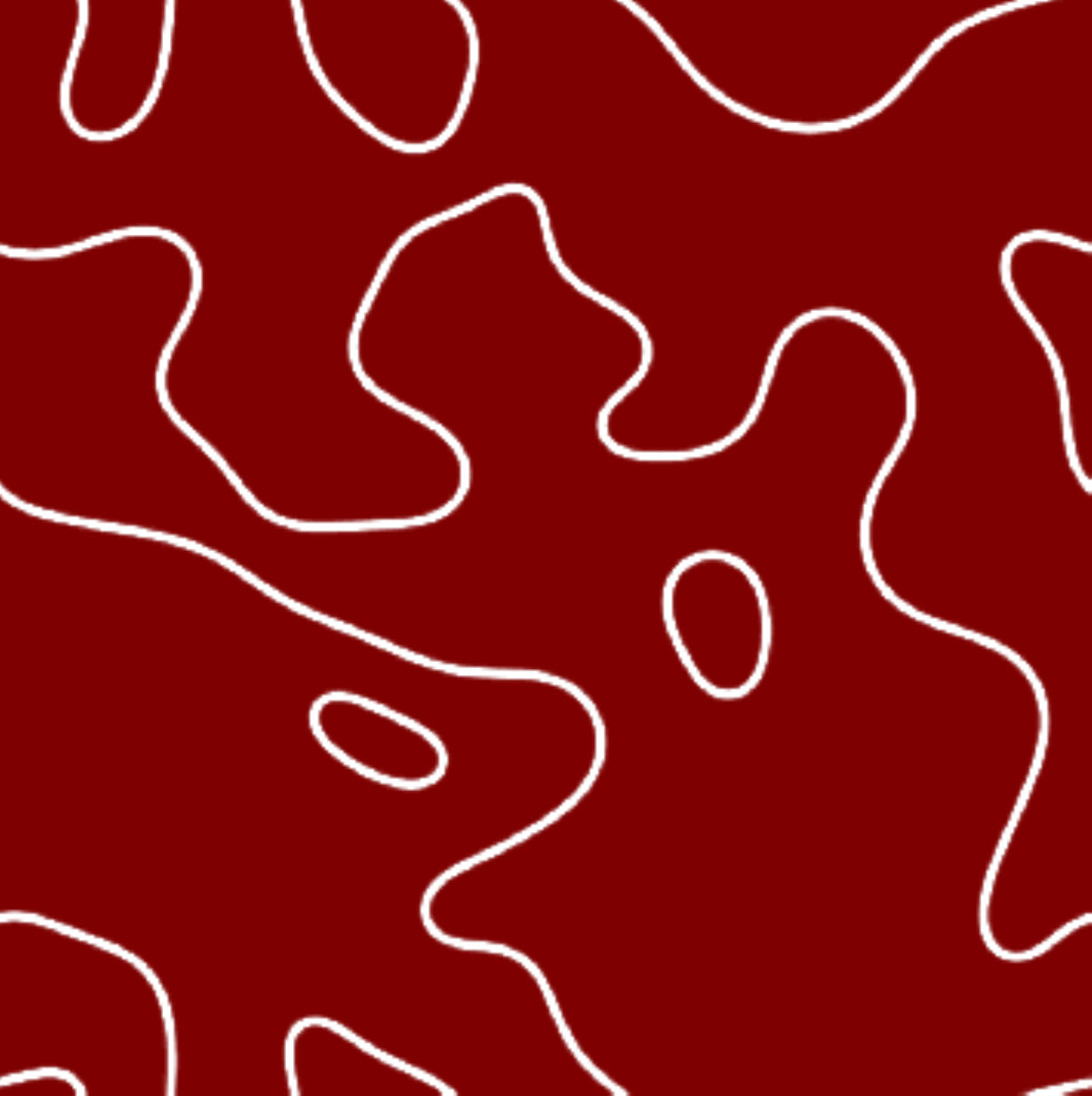}
	\includegraphics*[width=4.1cm]{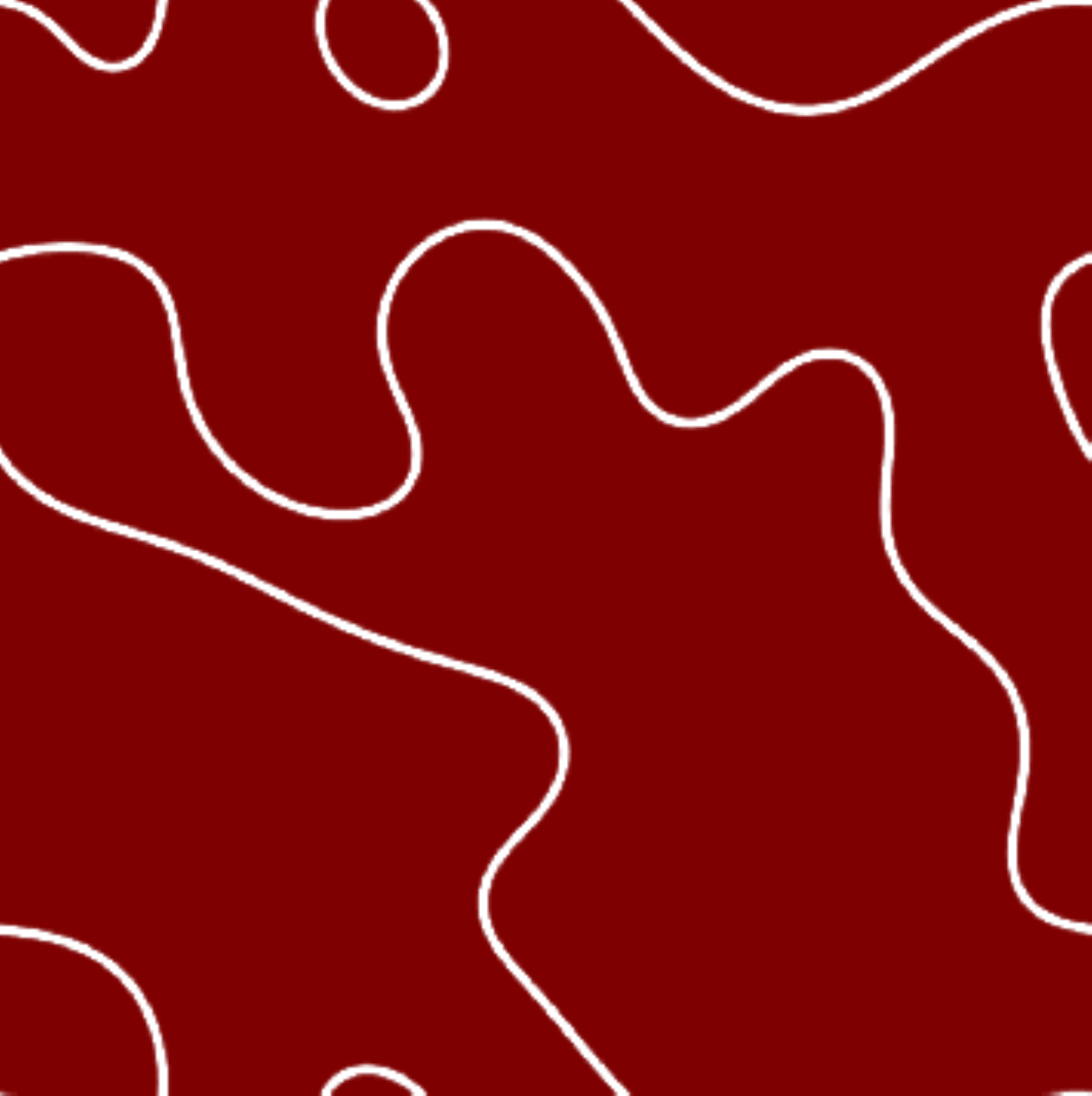}
	\includegraphics*[width=4.1cm]{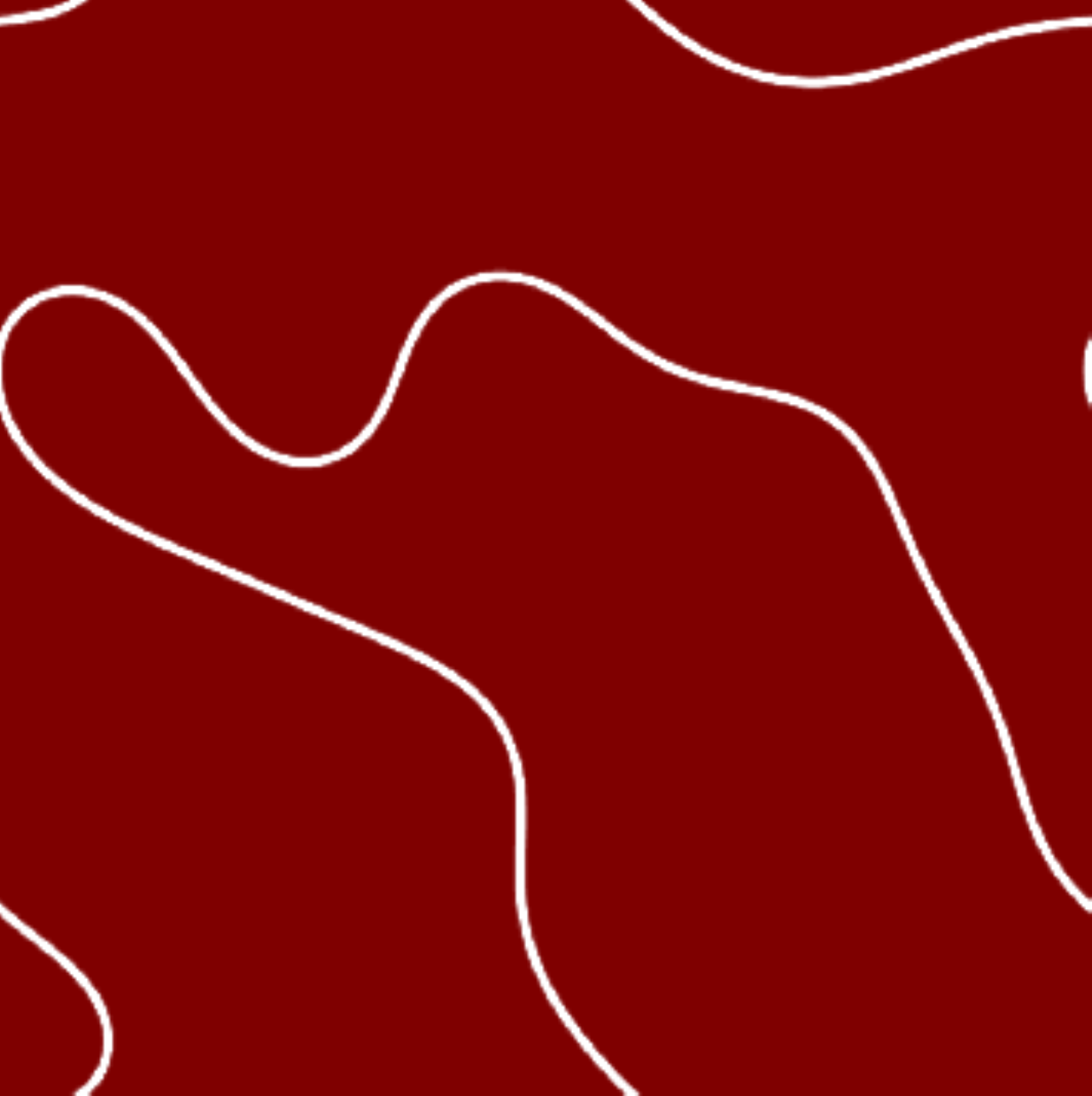}
	\caption{(Color online). The interface network of the snapshots shown in Fig.~1. Note that the white lines represent the spatial dispositions where $\phi_0$ departs from zero.}
	\label{fig3}
\end{figure}

We consider the scalar fields $\phi_1$ and $\phi_2$  which describe the densities of two competing species $1$ and $2$. In addition, we assume another scalar field $\phi_0$ that represents the number density of empty sites
created by the competition interactions between individuals of different species.
The individuals can also move on the network and reproduce. For simplicity, we shall consider that the interactions of both species
are quantified with same diffusion, reproduction and competition parameters, $D$, $r$ and $p$. 

The dynamics of the scalar densities is given by the mean field differential equations
\bq
\dot{\phi_1} &=& D\nabla^2\phi_1+r\phi_0\phi_1-p\phi_1\phi_2, \label{eq.phi1}\\
\dot{\phi_2} &=& D\nabla^2\phi_2+r\phi_0\phi_2-p\phi_1\phi_2, \label{eq.phi2}
\eq
where the dot represents a time derivative and $\nabla^2$ is the Laplacean operator. Note that the fields are constrained by
$\phi_0 + \phi_1 + \phi_2 =1$.

We solved numerically the mean field equations in two dimensional networks by starting with random
initial conditions, where at each grid
point a species s was chosen at random. In other words, we set $\phi_0=0$ and considered that
$\phi_i = 1$ if $i = \kappa$ and $\phi_i = 0$ if $i \neq\kappa$ at each grid point.

The time unit, one generation, is defined as the number of step times equals to the total number of grid points. 

Figure 1 shows four snapshots of a $1000^2$ lattice. 
The equations were implemented with $p=8.0$, $r=1.0$ and $D=1.0$, and the snapshots were captured after $1000$, $2000$, $4000$ and $8000$ generations.
The numerical results show that spatial domains with single species are formed. In fact, after the initial well mixed configuration, the system goes to a state where a low death rate is present, since the individuals search for their self-preservation in groups of their 
same species. 

The colors orange and green represent domains inhabited by individuals of species $1$ and $2$, respectively. In other words, in these regions $\phi_1$ and $\phi_2$ are equal to the unity, respectively (see also video in Ref. ~\cite{video1-2D}).  
These spatial patterns show that individuals of same species join each other for sharing regions on the lattice. 

The domains are bounded by interfaces mostly occupied by vacant sites. Figure 2 shows the interface networks of the snapshots of Figure 1, where it is highlighted the spatial distribution of number density of empty spaces $\phi_0$ on the 
lattice. Note that $\phi_1$ and $\phi_2$ do not vanish at the interfaces. This happens because whenever an individual is killed, the empty site can be filled by the offspring of neighbor individuals, that can move toward to the center of the interface before being caught by an enemy individual.

It has been shown in Ref. \cite{PhysRevE.86.031119} that the interfaces tend to straighten in order
that their total length decreases in time. As a result, some domains grow while other ones collapse as a direct consequence of the competition between the species. For a large number of generations, only one species invade the entire territory. We point out that the winner 
species is chosen randomly since it has been assumed that the interactions happens with the same rates for both strategies.

We also implement the differential equations assuming three dimensional networks as it is shown in Fig. 3. The results were obtained by assuming 
a $200^3$ grids and the snapshots were taken after $500$, $1000$, $2000$ and $4000$ generations. The red walls show the regions where $\phi_0$ 
departs from $0$, the borders of domains where attacks and counter attacks take place. The three-dimensional domains occupied by individuals of species 1 and 2 were left uncolored. The video in Ref. \cite{video1-3D} shows how the interface network evolves in time.

The dynamics of these interface networks is similar to that of domain wall networks in field theory models, where the characteristic length evolves as $L \propto t^{\frac12}$. This kind of curvature-driven network evolution appears in different scenarios in non-linear systems. Due to this similarity, we will further focus on describing the solitonic behavior of the internal structure of the interface. 

Finally we stress that the changes in the field $\phi_0$ through the interface happens only in its transversal direction separating domains 
with different species. In other words there no changes along the surfaces of the planar interfaces shown in Fig. 3.

\section{The single scalar field model} 
We now search for a field model that describes the topological aspects of the interfaces in systems with two competing strategies.
We point out that although the interfaces separating the domains appear in two and three dimensional spatial patterns, as it is shown in Fig. 2 and 3,
their profiles are defined in only one spatial dimension. To be more precise, the profile is the one dimensional density of empty space at their cross section. Their configuration is constant throughout the lattice and does not change in time, so that we shall focus on the
static distribution of the field by assuming the equations
\bq
\frac{d^2\,\phi_1}{dx^2}+a\phi_0\phi_1-b\phi_1\phi_2\,&=&0, \label{eq.st-phi1}\\
\frac{d^2\,\phi_2}{dx^2}+a\phi_0\phi_2-b\phi_1\phi_2\,&=&0, \label{eq.st-phi2}
\eq
where $a=r/D$ and $b=p/D$.

In order to build a spontaneous symmetry breaking formalism, we consider a new scalar field defined by 
\be
\Phi=\sqrt{\frac{2\,H}{a}}\,\left(\phi_2-\phi_1\right),
\ee
where $H$ is the maximum value of $\phi_0$, which it is defined as the interface height. In other words, $H$ is the value assumed by $\phi_0$ 
at the center of the interface. 

Using this new scalar field Eqs. \ref{eq.st-phi1} and \ref{eq.st-phi2} can be replaced by a single equation given by 
\be
\frac{d^2\,\Phi}{dx^2} = - a\,\phi_0\Phi \label{eq.static},
\ee
which it is proved to be useful to describe the one-dimensional variation of the field across the interface.

In this formalism the energy density of the system is given by
\be
\mathcal{E} = \frac{1}{2}\left(\frac{d\Phi}{dx} \right)^2+V(\Phi)\label{phizero},
\ee
where $V(\Phi)$ is the potential. This expression can be reduced to $\mathcal{E}=2\,V(\Phi)$ for stable solutions of the equations of motion
with $Z_2$ symmetry, that indicates the presence of two states of minimum energy \cite{bolg,PhysRevLett.35.760}.
We stress that the term energy here is used only for associating the competition system with the spontaneous symmetry breaking formalism of field theory.

\begin{figure}
	\centering
	\includegraphics*[width=4.2cm]{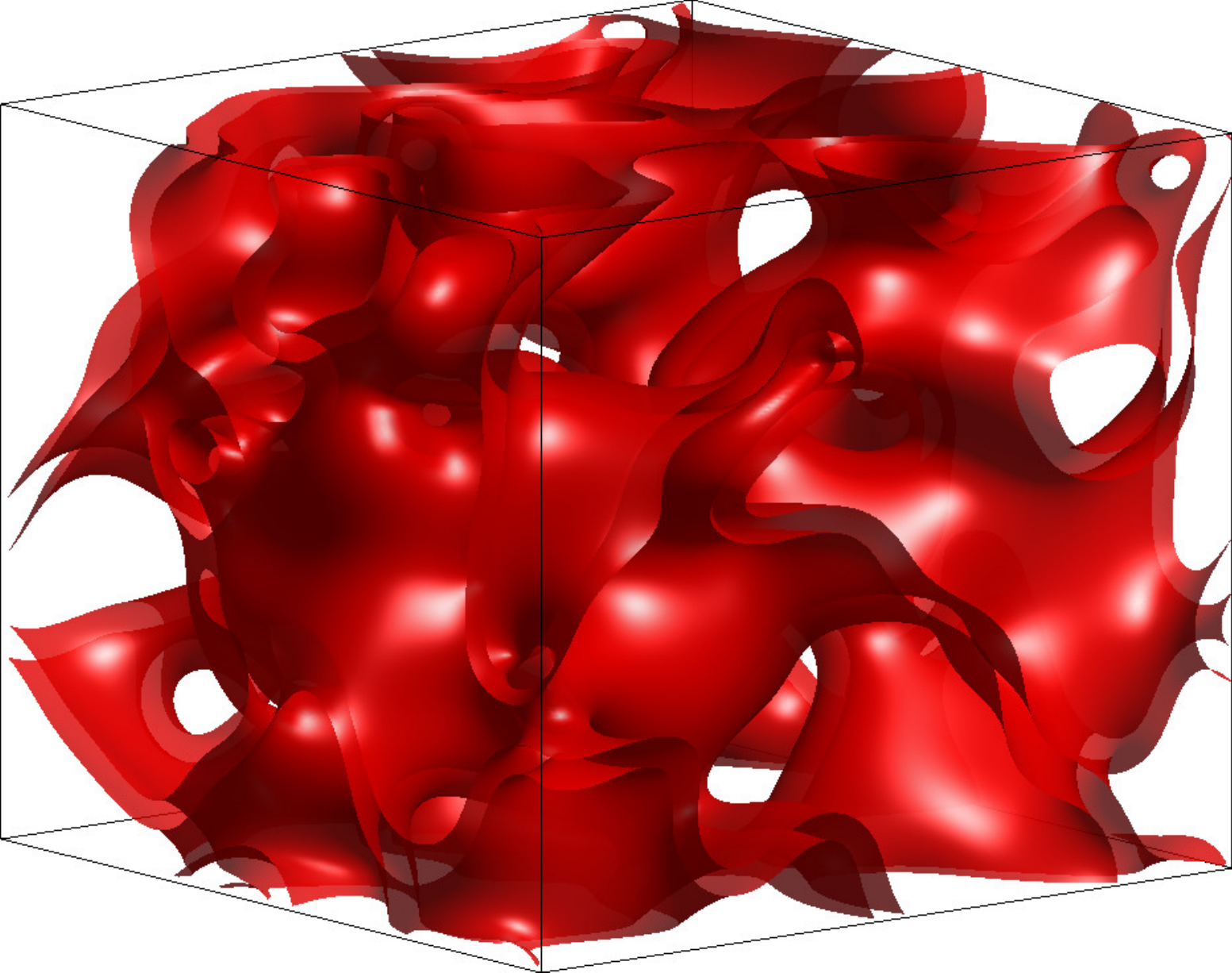}
	\includegraphics*[width=4.2cm]{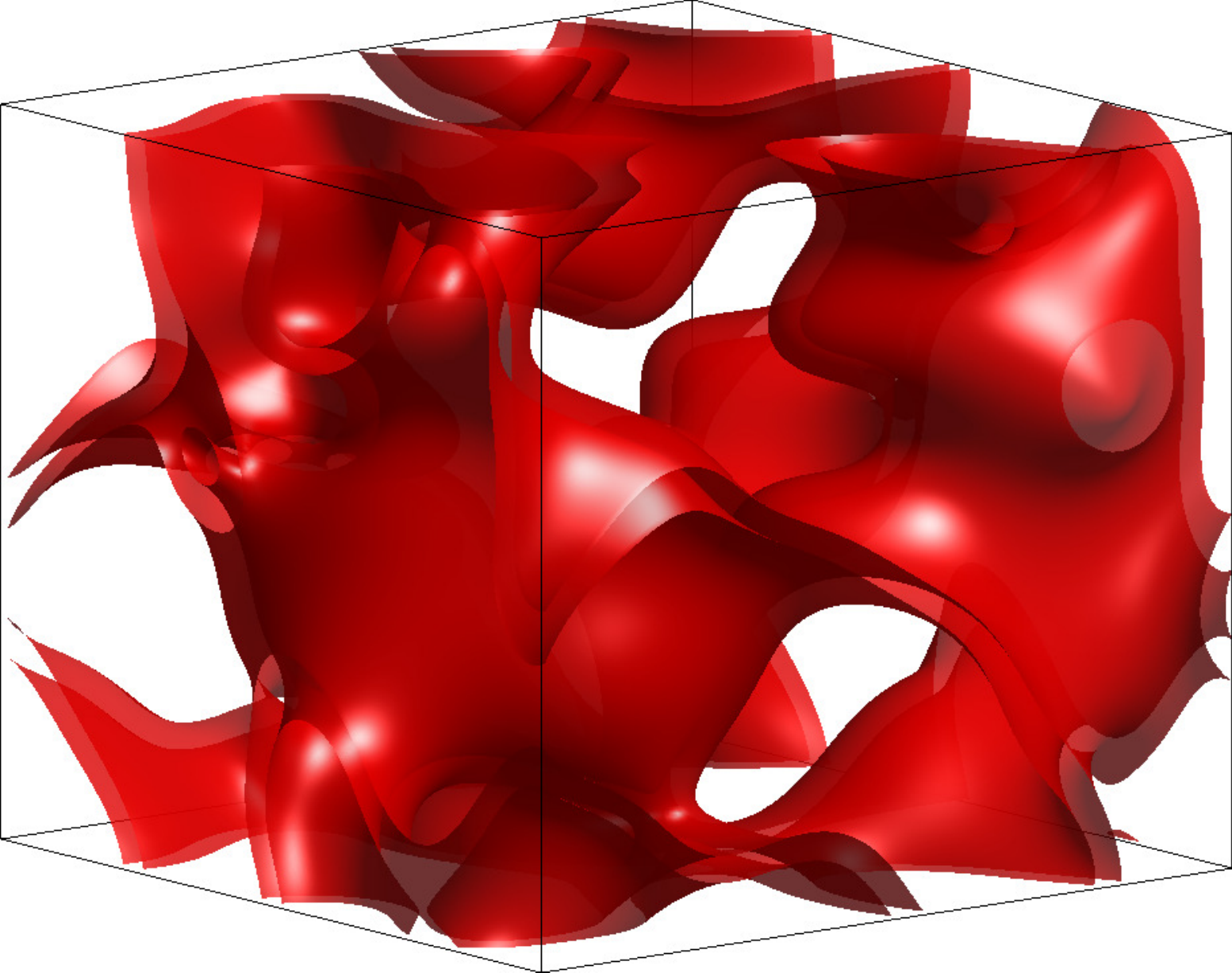}
	\includegraphics*[width=4.2cm]{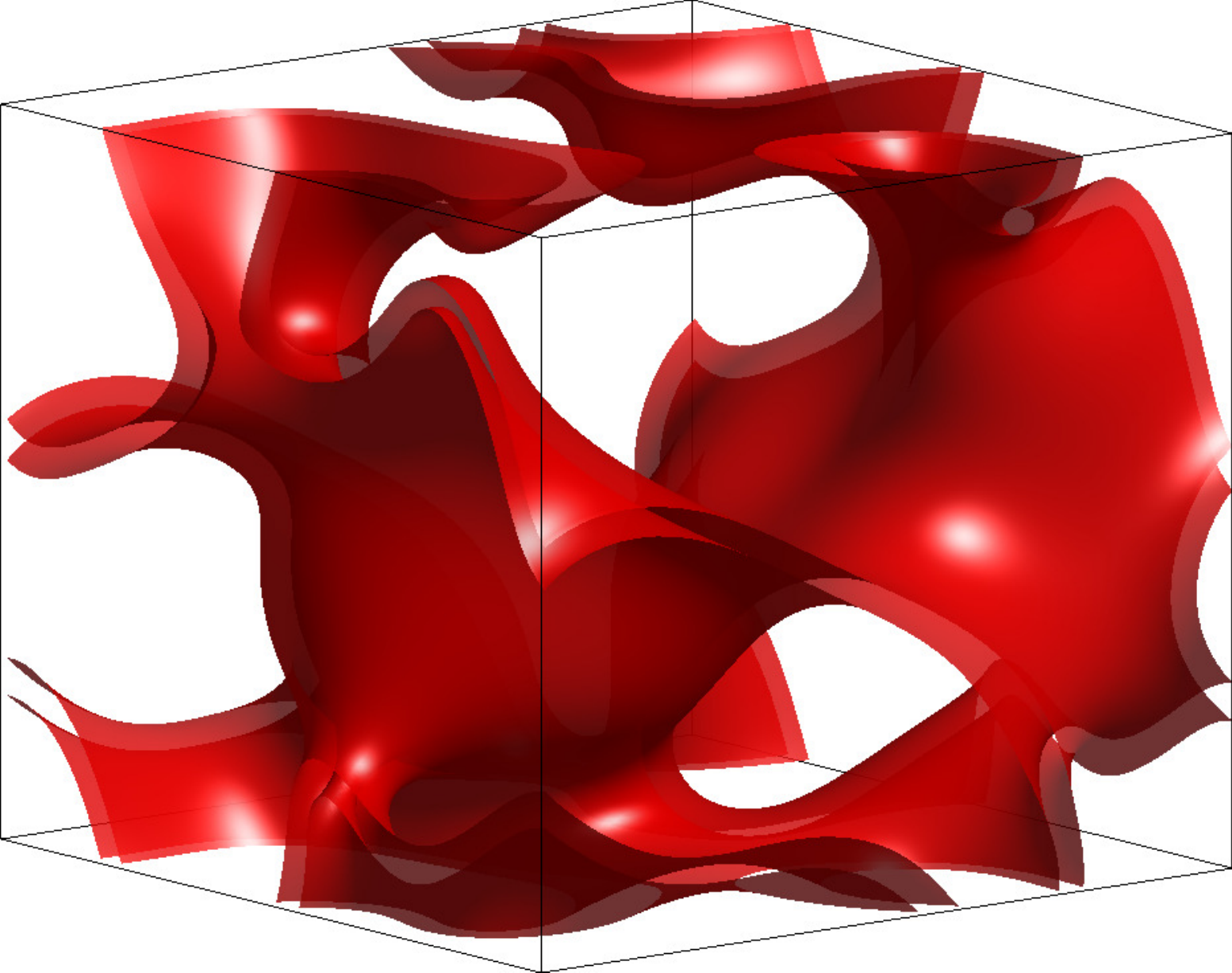}
	\includegraphics*[width=4.2cm]{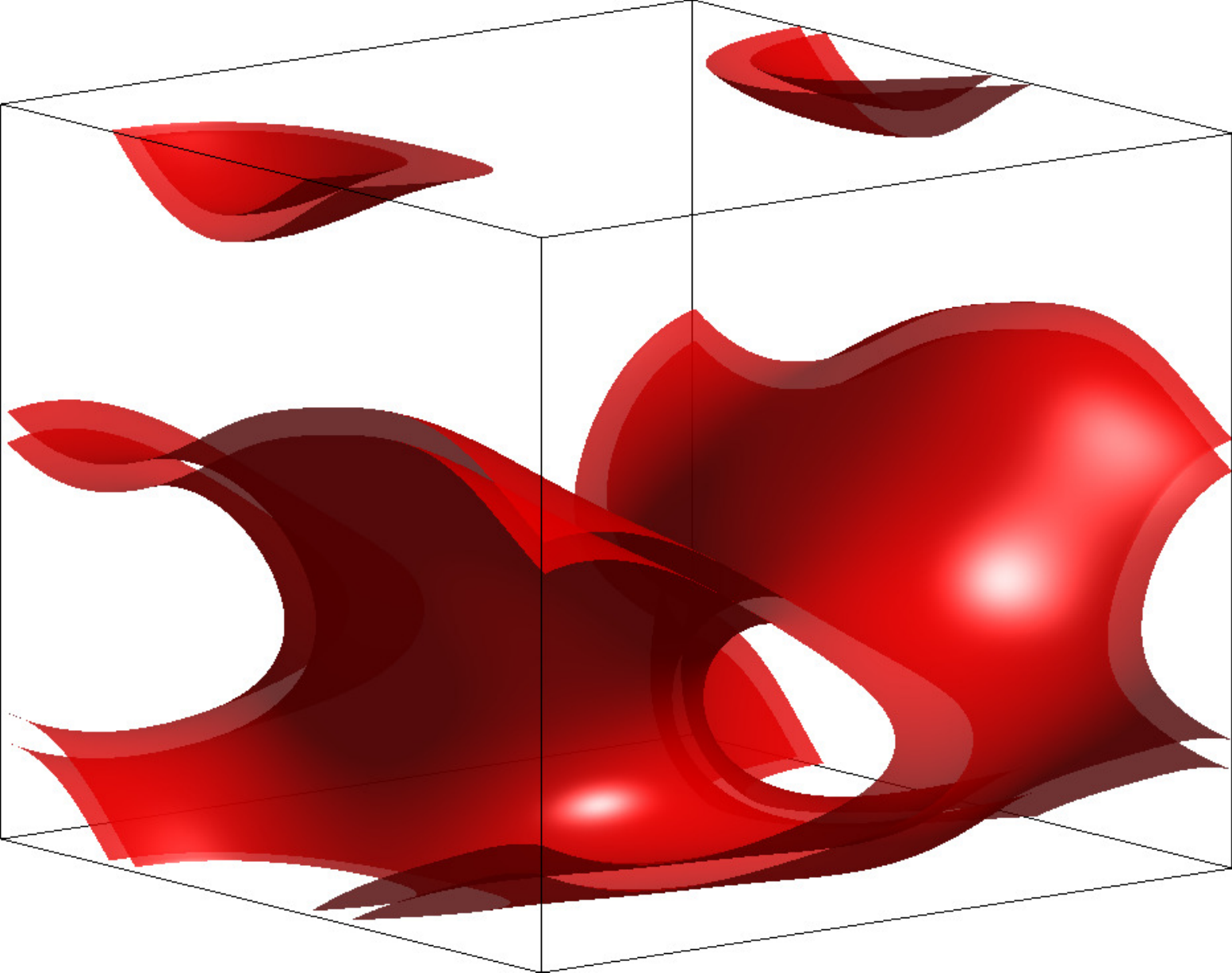}
	\caption{(Color online). Snapshots of a three dimensional network $200^3$ implementation of the mean field equations, taken after
	500, 1000, 2000 and 40000 generations. The spatial domains composed by individuals of single species, where $\phi_0 =0$, is left uncolored. 
	The red walls show the interfaces on the borders of the domains, mostly composed by empty space.}
	\label{fig3}
\end{figure}

Applying this mathematical framework to the competing species, the $Z_2$ symmetry leads to appearance of two different domains of single species when
the system undergoes to one of the potential minima. In other words, whenever one spatial region is filled by a single species (orange and green domains in Fig. 1), the density of empty spaces is reduced to zero, representing one minimum energy state of the system.  In fact, inside the domains the number of empty sites is null since no competition between individuals of same species is considered.
 
On the other hand, in the boundaries of two different domains, competing individuals attack each other, forming interfaces of empty sites. The density of such vacancies has maximum value at the center of the interface, which we have defined as the interface height.

\begin{figure}
    \center
    \includegraphics[width=8cm]{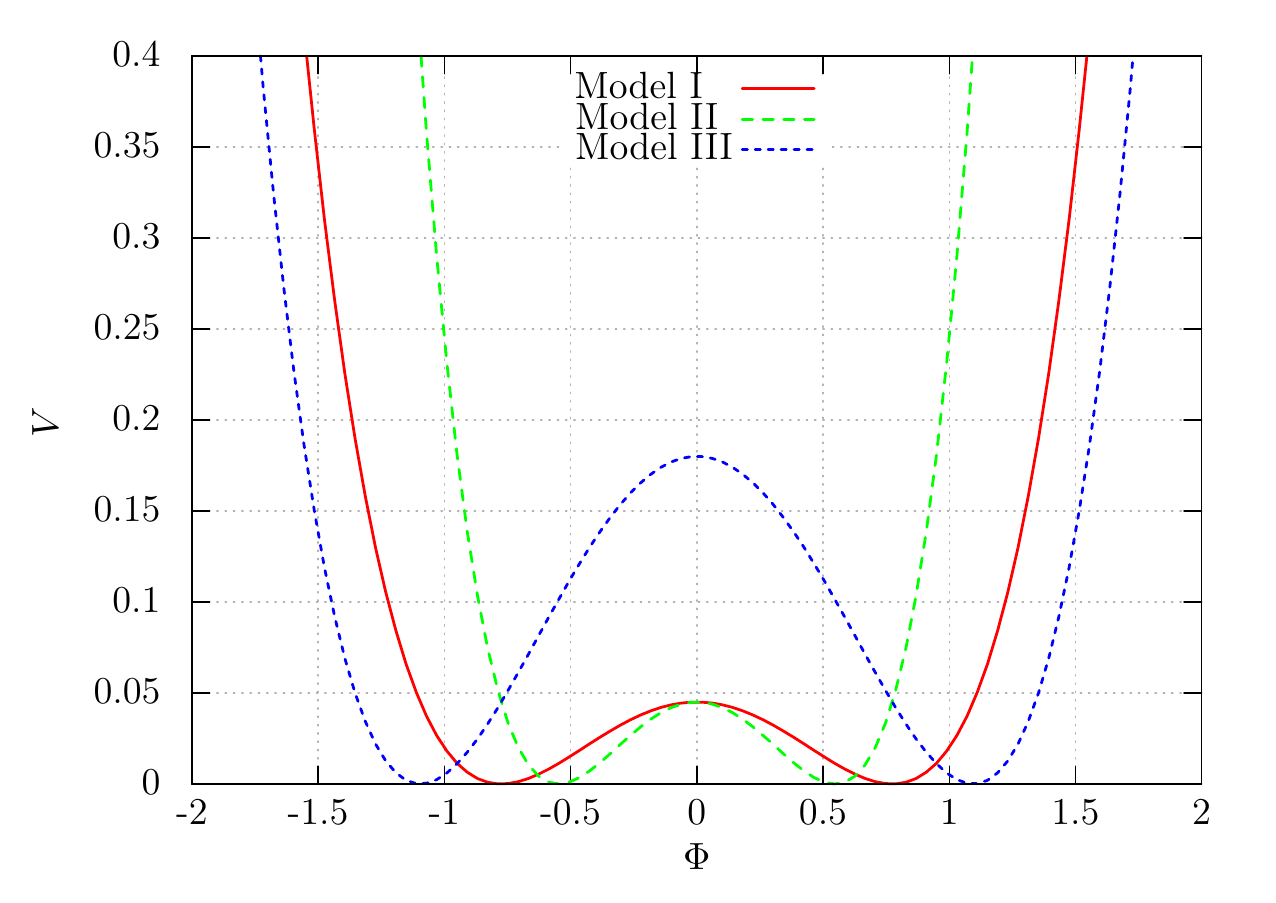}
    \label{graf0}
    \caption{The potential (Eq.~\ref{potential}) for Models I, II and III. The potential height and width are controlled by the parameters $H$ and $a$, as it is indicated in the figure.}
\end{figure}
The interface profile is found by solving the equation of motion
\be
\frac{d^2\,\Phi}{dx^2} = \frac{d\,V}{d\,\Phi} \label{eq.motion-static}.
\ee
Using the same parameters of Eqs. 1 and 2, we propose the modeling of the interface properties by assuming the potential
\be
V(\Phi)= \frac12\, \left(H-\frac12 a\,\Phi^2\right)^2.
\label{potential}
\ee
The domains occupied by species $1$ or $2$ arise when the system undergoes a phase transition, from $\Phi=0$ to $\Phi = -\sqrt{2\,H/a}$ or $\Phi = \sqrt{2\,H/a}$, respectively.

Figure 4 shows the potential for three different sets of parameters, that we shall refer as Model I ($H=0.30$, $a=1.0$),
Model II ($H=0.30$, $a=2.0$) and Model III ($H=0.60$, $a=1.0$). Note that the height and width of the potential are controlled by the model parameters, which are directly related to the interface properties.

We highlight that $H$ is a function of the parameters $a$ and $b$, which makes the topological modeling of the competition system dependent on the 
same parameters which appear in the mean field equations. We verified how $H$ behaves by carrying out a series of one-dimensional numerical implementations of Eqs. 1 and 2. Initially we assume that $\phi_1=1$ and $\phi_2=1$, for $x<0$ and $x>0$, respectively. After a few number of time steps, competition interactions give rises a stable central interface with constant height and width. The interface is centered at $x=0$, where the fields assume
the values $\{\phi_0=H,\phi_1=\phi_2=(1-H)/2\}$. 

The interface height is shown in Fig. 5 for a wide range of parameters $a$ and $b$ 
Note that the larger $b$ (or smaller $a$) the higher the interface. In other words, the number of vacancies created by attacks between the competing species is increased. On the contrary, for small $b$ (or large $a$) the species mostly move and reproduce filling the vacancies and decreasing $\phi_0$. 

We point out that the potential in Eq. 9 was introduced by assuming that
the density of empty spaces and the energy density in the theoretical framework are related by $\phi_0 = \sqrt{\mathcal{E}}$.
Although we have written the number density of empty spaces as a function of the energy density of the scalar field system, it
does not mean that these quantities have the same physical meaning. In fact, by matching both formalism, this function allows the investigation
of the interface networks by using the theoretical formalism widely considered in nonlinear science. 

\begin{figure}
    \center
    \includegraphics[width=8cm]{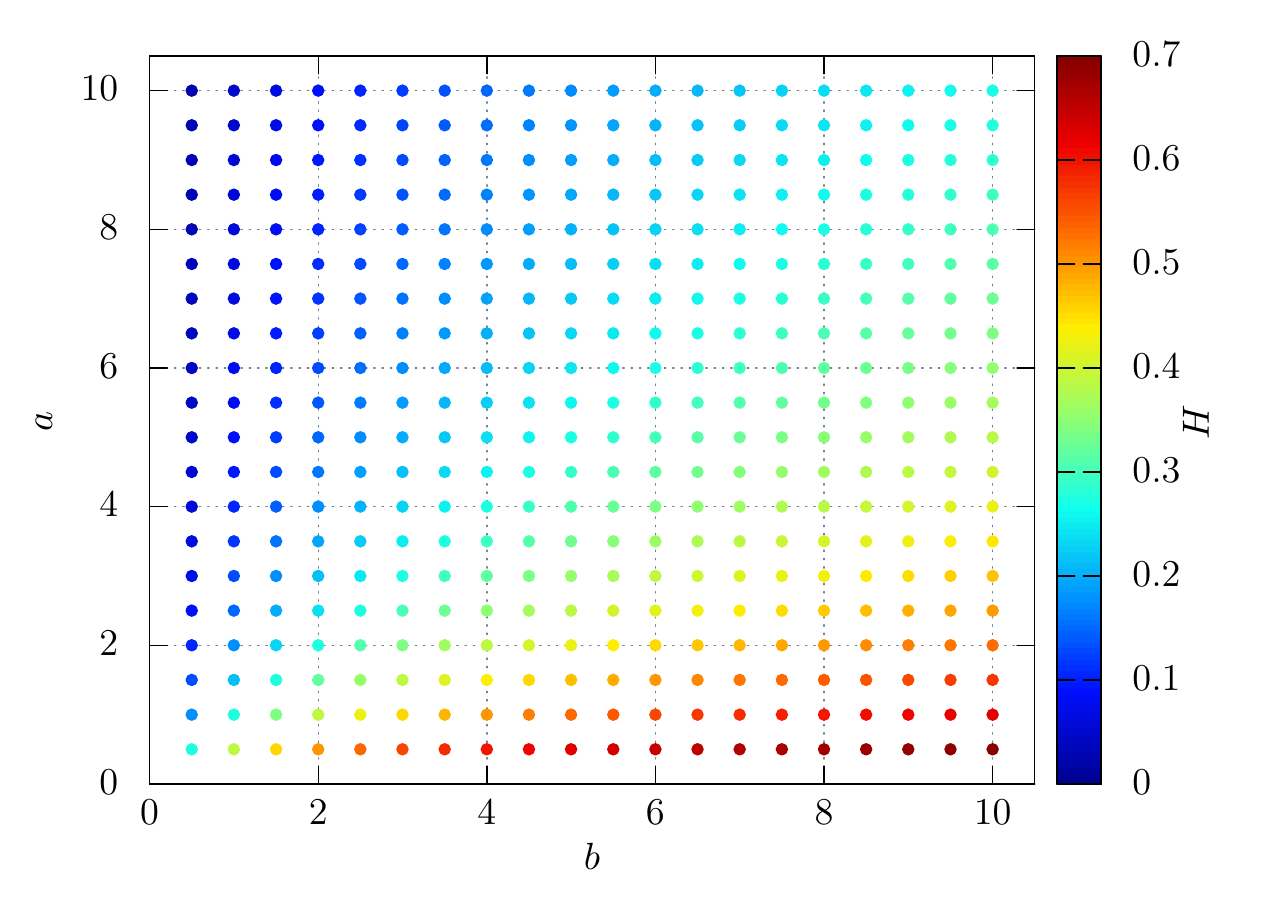}
    \label{fig5}
\caption{The interface height as a function of $a$ and $b$. The results were obtained by means of 400 one-dimensional numerical implementation of Eqs. 1
 and 2. The colors of the dots represent $H$ for each set of parameters.}
\end{figure}

\section{The interface profile}

The field $\Phi$ is then found by solving the equation of motion for Eq.~ 9, which gives
\be
\Phi(x)=\pm \sqrt{\frac{2\,H}{a}}\,\tanh \left(\sqrt{\frac{a\,H}{2}}\,x \right). \label{solitonic}
\ee   
These solutions are named solitons due to their topological properties, since they connect the potential minima. 
Indeed the positive solution represents $\phi_1$ going from 1 to 0 through the interface, while the negative solitonic solutions indicates the change from spatial domains of species 2 to species 1.

Figure 6 show the positive solutions $\Phi$ for Models I, II and III.
Note that $\Phi$ goes to different values asymptotically, which 
ensures the topological features. The parameters of the model controls such asymptotic behavior, since $H$ and $a$ determines the states of 
minimum energy and how fast the solutions reach them. 

Therefore, the analytical function of the interface profile is given by 
\be
\phi_0(x)\,=\,H-H\,\tanh^2\left(\sqrt{\frac{a\,H}{2}}\,x \right),
\ee
whose center is located at $x=0$, so that $\phi_0(x=0)=H$. The solid lines in Figure 5 show the interface profile for the Models I, II and III.
\begin{figure}
    \center
    \includegraphics[width=8cm]{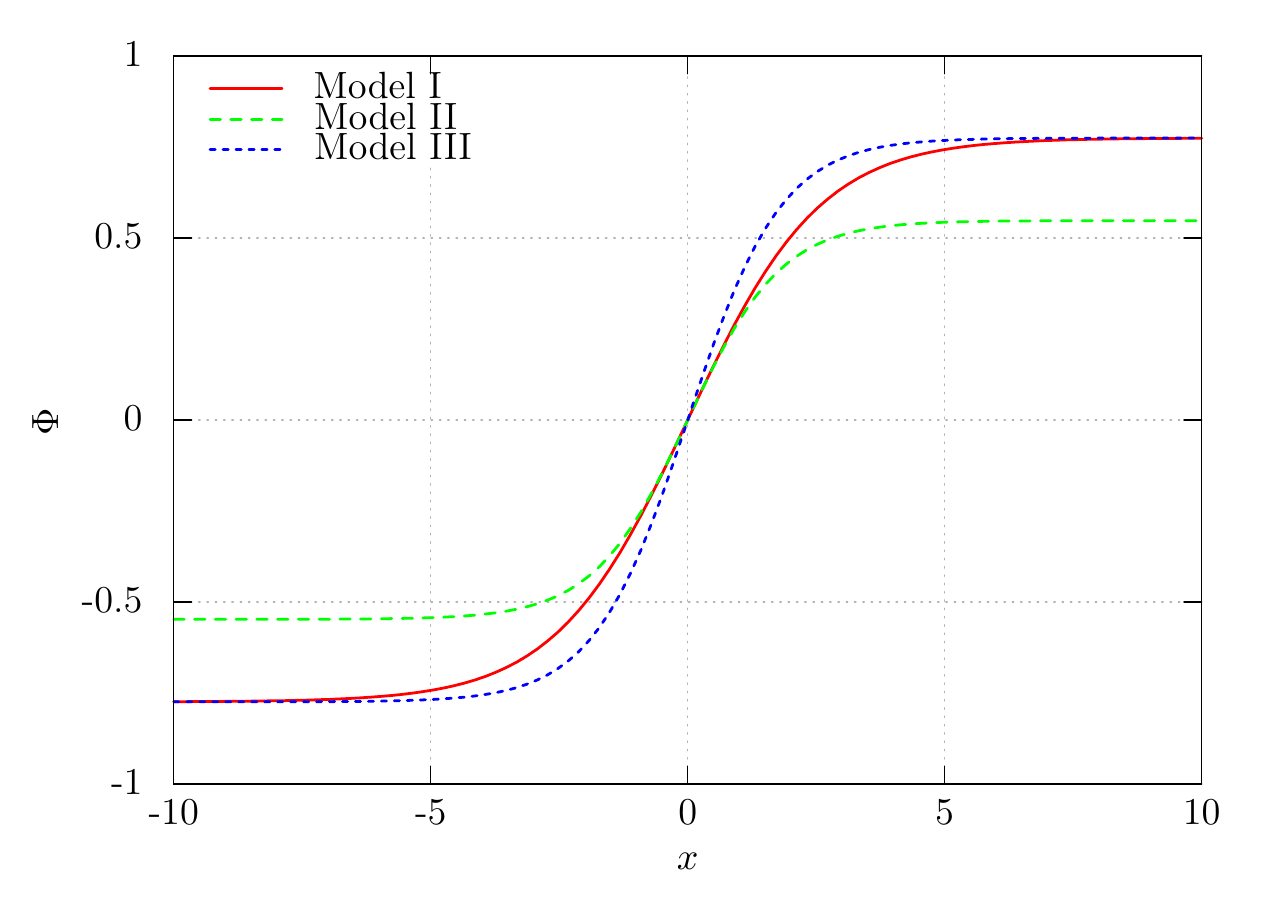}
    \label{graf1}
    \caption{The positive solitonic solutions $\Phi(x)$ for the potentials plotted in Fig. 2. Note that as $\Phi$ changes between the potential minima, the spatial distribution goes from domains of species 1 to domains of species 2.}
	\label{s}
\end{figure}

This theoretical approach gives that the interface width 
is a function of the parameters given by 
\be
\delta \simeq \sqrt{\frac2{H\,a}}.
\ee

In addition, the interface profile can be characterized by different topological properties of the solitonic description. For example, the effective energy and
the topological charge, given respectively by
\be
E_{eff} = \int\,dx\,\phi_0^2\,=\,\frac43\,\sqrt{\frac{2\,H^3}{a}}
\ee
and
\be
Q_{eff} = \int\,dx\,\phi_0\,=\,2\,\sqrt{\frac{2\,H}{a}}
\label{tcharge}
\ee
are alternative ways to compute the role of the interfaces in the system. 
Moreover, as functions directly and inversely proportional to $H$ and $a$, respectively,
They may be used to quantify the presence of empty space per unit length of the interfaces. In fact, the effective energy and topological charge
diverge for $a=0$, which means that no offspring is generated. In this situation the entire network would be composed by empty spaces, and consequently, the interface would have infinite width and energy (very large width and energy, taking into account the finiteness of the network). On the contrary, if only reproduction is present ($a$ very large) the interface is eliminated, that is $\delta=E_{eff}=0$.

In summary, by means of the one dimensional function $\Phi(x)$, all topological aspects of the interface profile connecting both sides of the battlefront are found. Immersed in higher spatial dimensions, the solitonic solutions describe the linear and planar interfaces which appear in Fig. 2 and 3. 
These configurations are stable against small temporal perturbations, as one can be verified by substituting the perturbed function
$\Phi(x,t)=\Phi(x)+\sum_{k}\,\eta_{k}(x)\,\cos(\omega_{k}\,t)$ in the equation of motion. The stability equation 
\be
-\frac{d^2\,\eta_{k}}{dx^2}+a\,H\left[3\,\tanh^2\left(\sqrt{\frac{a\,H}{2}}\,x\right)-1\right]\eta_{k}=\omega_{k}^2\,\eta_{k}
\ee

has solutions that leads to $\omega_{k}^2 \geq 0$. This implies that despite the linear fluctuation the solution remains stable.
The stability of domain wall profiles has been studied in systems with single or multiple scalar fields in Physics (see Refs. 
\cite{landau,vilenki,jackiw,PhysRevE.54.2943,0305-4470-30-23-015}, for example). Applied to population dynamics, this
property provides a proof that the interface networks are stable spatial pattern configurations in systems with competing species.

\section{Discussion}
\begin{figure}
    \center
    \includegraphics[width=8cm]{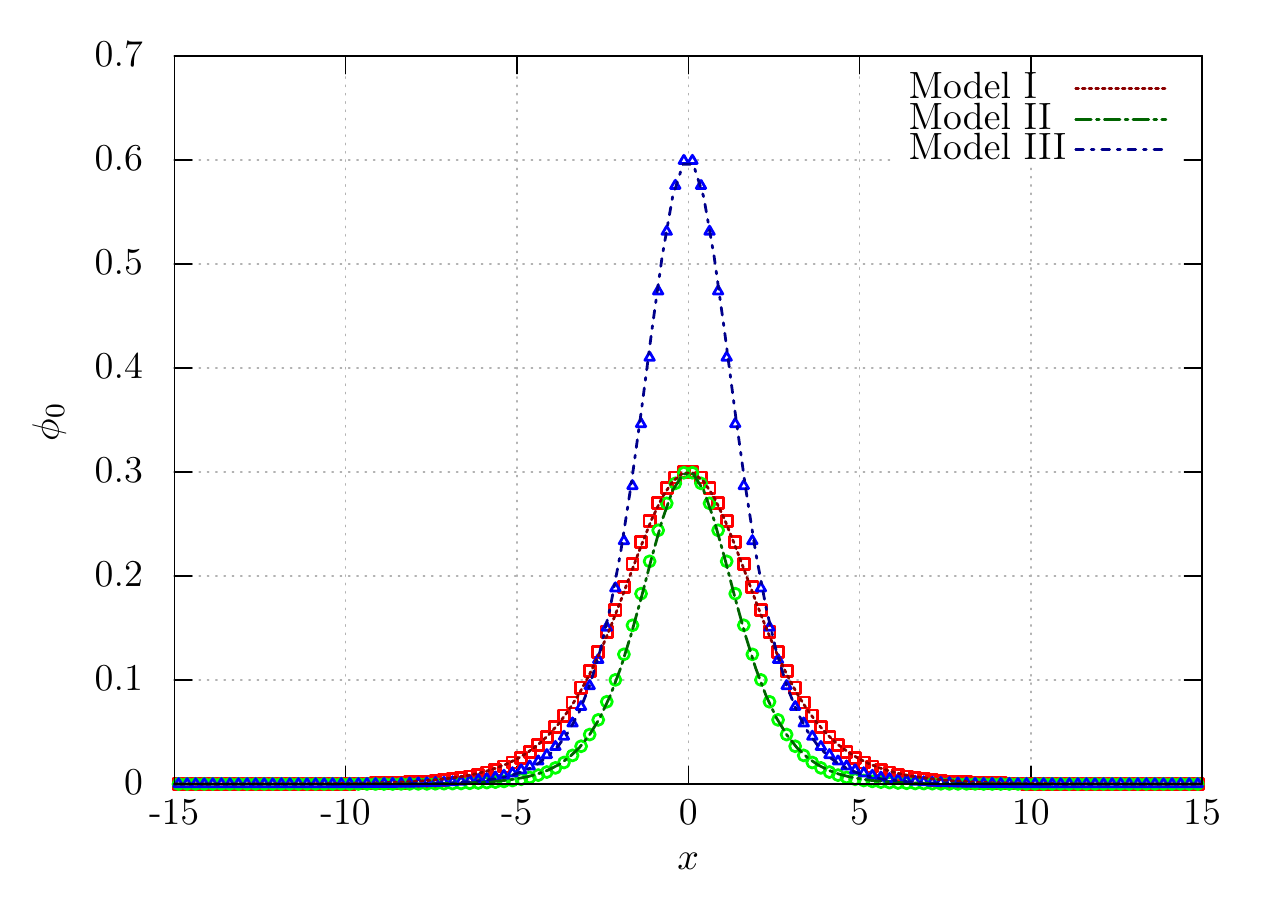}
    \label{graf2}
    \caption{Interface profiles for the Models I, II and III. The dashed lines represent the theoretical functions, whereas the 
    red squares, green circles and blue triangles were obtained by means of numerical implementations of Eqs. 1 and 2.}  
\end{figure}

In order to verify the accuracy of our theoretical approach to fit the numerical results obtained by the implementations of the mean field equations, we plotted them together in Fig. 7.
\begin{figure}
    \center
    \includegraphics[width=8cm]{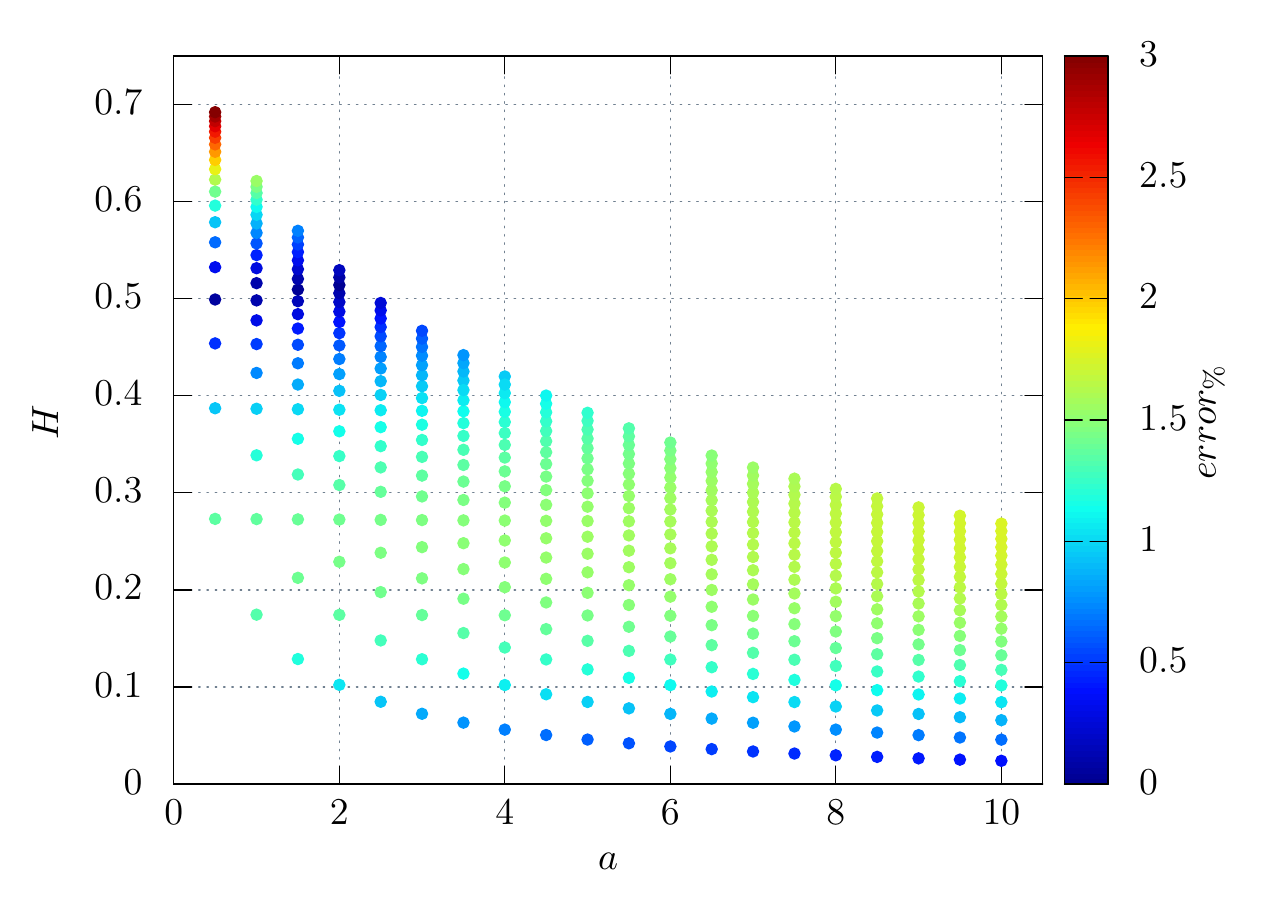}
    \label{fig5}
\caption{Comparison of theoretical topological charge with the numerical implementation of Eqs. 1 and 2.
Note the colors of the dots show that the relative error is very small for a wide range of parameters $a$ and $H$.}
\end{figure}
The red, green and blue dashed lines were show the analytical function for the profile interface for the Model I, II and III, respectively. 
Assuming $b=1.2$, $b=2.4$ and $b=8.0$ in Eqs.~\ref{eq.st-phi1} and \ref{eq.st-phi2}, 
we recovered the same models numerically. The results are represented respectively by red squares, green circles and blue triangles. 
The agreement between both plots allows us to conclude that
our theoretical function fits well the interface profile.

Moreover, by using the solitonic description, we compute the topological properties of the interface for the Models I ($\delta=2.58$, $E_{eff}=0.31$, $Q_{eff}=1.54$), II ($\delta=2.58$, $E_{eff}=0.62$, $Q_{eff}=1.09$) and III ($\delta=1.82$, $E_{eff}=0.87$, $Q_{eff}=2.19$). These results show that for fixed $a$, the interface height $H$ is increased when a higher $b$ is assumed. This is a consequence of intensifying competition on the battlefronts. On the other hand, for fixed $H$, as $a$ increases the interface width decreases. Therefore as $a=r/D$, the narrowing of the interface is achieved either by taking a higher reproduction parameter $r$ (less number density of empty spaces on the borders of the domains) or a lower diffusion parameter $D$ (lower mobility prevents individuals of going towards the opposite territory).   

Finally, we made quantitative comparison between the theoretical properties of the interfaces and the results provided by the numerical implementation
of Eqs. 1 and 2. We calculated the topological charge by integrating $\phi_0$ and compared with Eq.~\ref{tcharge}.
Figure 8 shows the relative errors on the analytical approach as a function of $a$ and $H$. Note that 
the relative is small for a wide range of parameters.

Based on these results we conclude that not only does agree our theoretical description with the numerical implementation of the mean field equations, but it also provides a accurate way to quantify the physical properties of the interface.

\section{Conclusions}

The interface profiles has topological properties which are well described by solitonic solutions of a single scalar field model.
The approach presented in this paper has a potential with $Z_2$ symmetry. The spontaneous symmetry breaking provides the formation of two types of domains, representing spatial regions with individuals of distinct species. 
The analytical solution of the equation of motion yields to all topological properties of the interface profiles and strongly agree with the numerical
results provided by the implementation of the equations of motion.

Another advantage of the solitonic description is the stability of solutions of the equations of motion against small temporal perturbations. 
Furthermore the stability of the spatial pattern networks plays an important role in the understanding of population dynamics, since it allows the prediction of the way populations
evolve and how extinction takes place.

The solitonic description of interface profiles can be generalized to models with a larger number $N$ of species, either forming $N/n$ partnerships of $n$ species or competing each other. In this generalized models the interfaces might join each other in $Y$-type or $N$-type junctions, depending on the effective energy of the different interfaces present in the network. Our model can be used to investigate how the junctions are preferred according to the intrinsic effective energy and topological charge of different interfaces.

Finally we point out that the topological description used to investigate interfaces with discrete symmetry can be extended to 
more complex spatial patterns. The same spontaneous symmetry breaking process is largely used to study how strings networks arise and evolve in 
Cosmology and Condensed Matter. We claim that the same theoretical framework can be applied to describe the string networks in 
Lotka-Volterra competition scenarios.

\begin{acknowledgments}

We thank CAPES, CNPQ/Fapern and IBED-Universiteit van Amsterdam for financial support. 

\end{acknowledgments}

\bibliography{z2-competition}

\end{document}